\documentclass[twocolumn,amsmath,superscriptaddress,amssymb,nopacs,nofootinbib,prl]{revtex4-2}

\usepackage[usenames,dvips]{color}
\usepackage{graphicx}
\usepackage{dcolumn}
\usepackage{bm}
\usepackage{mathtools}
\usepackage{epstopdf}
\usepackage{esint}
\usepackage{xcolor}
\usepackage{dsfont}
\usepackage[outline]{contour}
\usepackage{nicefrac}

\usepackage[colorlinks=true,citecolor=magenta,linkcolor=magenta,urlcolor=magenta]{hyperref}

\newcommand{\bea}{\begin{eqnarray}}
\newcommand{\eea}{\end{eqnarray}}
\newcommand{\bt}{\textbf}

\newcommand{\ph}{\phantom{.}}

\newcommand{\noi}{\noindent}
\newcommand{\no}{\nonumber}

\newcommand{\PK}[1]{\textcolor{black}{#1}}

\begin{document}
\def\v#1{{\bf #1}}

\title{Topological Superfluid Responses of Superconducting Dirac Semimetals}

\author{Jun-Ang Wang}
\email{wangjunang@itp.ac.cn}
\affiliation{CAS Key Laboratory of Theoretical Physics, Institute of Theoretical Physics, Chinese Academy of Sciences, Beijing 100190, China}
\affiliation{School of Physical Sciences, University of Chinese Academy of Sciences, Beijing 100049, China}

\author{Mohamed Assili}
\email{m.assili@itp.ac.cn}
\affiliation{CAS Key Laboratory of Theoretical Physics, Institute of Theoretical Physics, Chinese Academy of Sciences, Beijing 100190, China}

\author{Panagiotis Kotetes}
\email{kotetes@itp.ac.cn}
\affiliation{CAS Key Laboratory of Theoretical Physics, Institute of Theoretical Physics, Chinese Academy of Sciences, Beijing 100190, China}

\vskip 1cm
\begin{abstract}
We demonstrate that topological constraints do not only dictate the geometric part of the superfluid stiffness, but can also govern the \textit{total} superfluid stiffness. By introducing a general adia\-ba\-tic approach for superfluid responses, we showcase such a possibility by proving that the stiffness of a superconducting Dirac cone in two dimensions (2D) is proportional to its to\-po\-lo\-gi\-cal charge. By relying on the emergent Lorentz invariance of Dirac electrons,  we unify the superfluid stiffness and quantum capacitance in these systems. Based on this connection, we further predict a topological origin for the quantum capacitance of a Josephson junction where 2D massless Dirac electrons are sandwiched between two conventional superconductors. We show that the topological responses persist upon effecting strain, are resilient against weak disorder, and can be experimentally controlled via a Zeeman field. Remarkably, the nonuniversal topological quantization of the two superfluid responses, yet implies the universal topological quantization of the admittance mo\-du\-lus of the superconduc\-ting Dirac system in units of conductance. The quantum admittance effect arises when embedding the superconducting Dirac system in an ac electrical circuit with a frequency tuned at the  absorption edge. These findings are in principle experimentally observable in graphene-superconductor hybrids.
\end{abstract}

\maketitle

The exploration of the interplay between topology and quantum geometry through superfluid responses of time-reversal inva\-riant superconductors (SCs)~\cite{Schrieffer} is currently in the spotlight~\cite{Torma,PeottaLieb,LongLiang,PeottaTormaBAB,RossiRev,Jonah}. Such a  pursuit got substantially boosted after expe\-ri\-men\-ts in magic angle twisted bilayer graphene (MATBG) provided evidences for a non\-va\-ni\-shing superfluid stiffness despite the almost flat energy dispersion which governs transport~\cite{Cao,Yankowitz}. This quite unusual result was subsequently understood in terms of lower bounds set by topological in\-va\-riants dictating the MATBG band structure~\cite{Ahn,Randeria,Pikulin,Julku,BABbounds,BABbounds2}. These bounds impose constraints on the so-called geometric contribution to the stiffness, which becomes relevant when supercon\-duc\-ti\-vi\-ty is harbored by flat bands. However, no topological constraints have been so far predicted for the \textit{total} superfluid stiffness. The potential discovery of systems whose total stiffness is equal or proportional to a topological inva\-riant promises to deepen our understanding of topological platforms and set the stage for novel applications.

In this Letter, we show that topological constraints dictate the total superfluid stiffness of superconducting Dirac semimetals (SDSs). Notably, a quantized total superfluid stiffness has already been theoretically predicted for superconducting graphene~\cite{KopninSoninPRL,KopninSoninPRB,KopninLayers,Peltonen}, which constitutes a prototypical SDS in its Dirac regime~\cite{Graphene}. However, the quantization itself, along with its to\-po\-lo\-gi\-cal origin and resulting implications have not yet been discussed. After Refs.~\cite{KopninSoninPRL,KopninSoninPRB,KopninLayers,Peltonen}, the superfluid stiffness of a supercon\-duc\-ting Dirac cone ($D_{\rm cone}$) reads at charge  neutrality as:
\begin{align}
D_{\rm cone}=
%\frac{e^2}{h}\frac{2\Delta}{\hbar}\equiv
\Delta/\pi\,,\label{eq:StiffnessUniversal}
\end{align}

\noi and becomes quantized in units of $\Delta\geq0$, which is the intrinsic or proximity-induced conventional pai\-ring gap felt by graphene~\cite{UchoaSeo,BreyFertig,Wakabayashi,covaci,bruno,Roy,V-Kauppila}. In the above, we set the reduced Planck constant $\hbar$ and the electric charge unit $e$ to unity.

\begin{figure}[t!]
\begin{center}
\includegraphics[width=1\columnwidth]{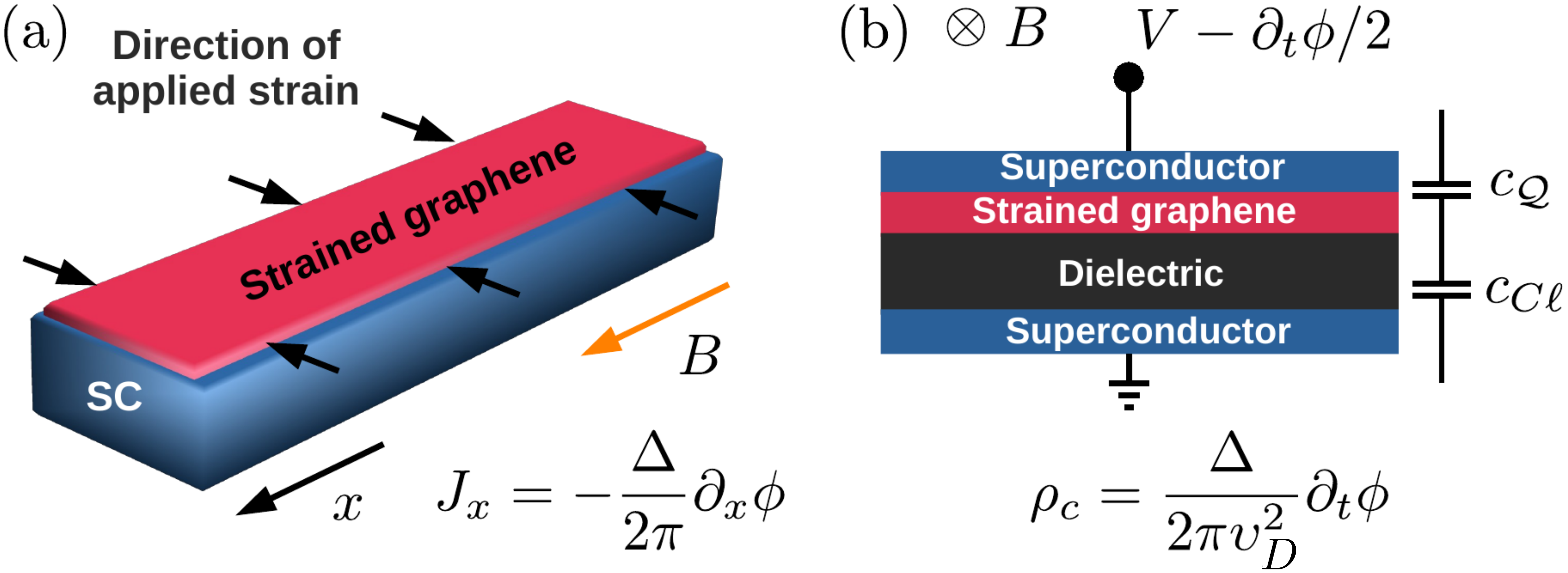}
\end{center}
\caption{Hybrids of strained graphene and superconductors (SCs). Each graphene Dirac cone leads to a quantized superfluid stiffness $D=-J_x/(\partial_x\phi/2)=\Delta/\pi$ and quantum ca\-pa\-ci\-tan\-ce $\PK{c_{\cal Q}=\rho/(\partial_t\phi/2)}=\Delta/\pi\upsilon_D^2$. A graphene cone sees a pairing gap $\Delta$ and Fermi velocity $\upsilon_D$. The quantized Dirac contributions can be disentangled by tuning strain, a Zeeman energy $B$, and/or the classical capacitance of the junction $c_{\PK{C\ell}}$.}
\label{fig:Figure1}
\end{figure}

\noi This quantization is rather puzzling as it cannot be explained by to\-po\-lo\-gi\-cal bounds of the type predicted for MATBG since, here, the dispersions are strongly non flat{\color{black}, and $D_{\rm cone}$ receives equal contributions from both interband (geometric) and intraband (conventional) parts~\cite{LongLiang}.}

Here, we resolve this conundrum by proving that $\pi D_{\rm cone}/\Delta$ is a topological invariant. This result is a consequence of non\-tri\-vial topology for a single Dirac cone in the presence of a phase which twists its mass, similar to the Jackiw-Rossi model~\cite{JRossi}. As we further explain in our companion work~\cite{WJAextended}, the quantization is also a manifestation of one-dimensional chiral anomaly~\cite{Semenoff,CA}. These properties, along with the arising Lorentz invariance, also render the quantum capacitance $c_{{\cal Q}}$ of a Josephson junction sandwiching a Dirac cone topological and equal to:
\begin{align}
c_{{\cal Q},{\rm cone}}
=D_{\rm cone}/\upsilon_D^2
\label{eq:CapacitanceUniversal}
\end{align}

\noi where $\upsilon_D$ is the group velocity of the Dirac dispersion.

Both effects enjoy topological robustness as long as the chiral symmetry of the Dirac Hamiltonian is preserved. To back this claim, we show that Eqs.~\eqref{eq:StiffnessUniversal} and~\eqref{eq:CapacitanceUniversal} persist upon adding nonuniform strain which leads to an energy spectrum that consists of re\-la\-ti\-vi\-stic pseudo-Landau le\-vels (pLLs)~\cite{NeekCovaciAmal,Neek-Amal,Salerno,Shuze,Nigge,Hsu,Jamotte,Tianyu22}. We show that for the pLL flat bands, $D$ and $c_{\cal Q}$ are purely of a geo\-me\-tric origin, and are solely carried by the zeroth pseudo-Landau le\-vels (0pLLs). {\color{black}Hence, we reveal that the topological quantization dictating the total superfluid stiffness of the SDS, further imposes the quantization for the geometric part of the stiffness when the Dirac bands flatten due to strain.}

We propose to experimentally detect these SDS effects in superconducting (un)strained graphene hybrids, such as the ones shown in Fig.~\ref{fig:Figure1}. The quantized contribution of the Dirac electrons can be disentangled by applying a magnetic field which couples to graphene electrons only through a Zeeman energy scale $B$. The Zeeman energy sets the occupancy of the energy levels of the SC. Hence, by ma\-ni\-pu\-la\-ting its strength $|B|$, one is in a position to controllably add or subtract the contribution of the Dirac part of the band structure. Most importantly, we bring forward that, at charge neutrality, the smoking gun signature of the topological effects predicted here is the observation of the universal topological quantization of the admittance modulus $Y_{\rm mod}$ of the SDS, when it is embedded in an ac electrical circuit. Specifically, when the ac frequency $\omega$ is tuned at the absorption edge $2\Delta$, $Y_{\rm mod}$ equals the number of Dirac cones in units of conductance.

We commence our analysis by unifying the superfluid stiffness tensor elements $D_{ij}$ and the Josephson quantum ca\-pa\-ci\-tan\-ce (JQC) $c_{\cal Q}$ in SDSs. The former quantities define the coefficients which link the charge current $\bm{J}(\bm{r})$ to the spatially dependent gauge invariant vector potential $\bm{A}(\bm{r})+\bm{\nabla}\phi(\bm{r})/2$, where $\bm{A}(\bm{r})$ denotes the electromagnetic vector potential and $\phi(\bm{r})$ is the super\-con\-duc\-ting phase. Therefore, we have the defining relations:
\begin{align}
J_i(\bm{r})=-D_{ij}\big[A_j(\bm{r})+\partial_j\phi(\bm{r})/2\big]\,,\label{eq:Current}
\end{align}

\noi where $i,j=x,y$ for a two-dimensional system. The JQC can be defined in an ana\-lo\-gous fa\-shion, through the excess charge density $\rho_c(t)$ induced in a Josephson junction due to the pre\-sen\-ce of the gauge invariant voltage $V(t)-\partial_t\phi(t)/2$. For a two-terminal Josephson junction as in Fig.~\ref{fig:Figure1}(b), $V(t)$ and $\phi(t)$ define the voltage and phase biases imposed across the junction. Therefore, we define $c_{\cal Q}$ as the response coefficient which satisfies the relation:
\begin{align}
\rho_c(t)=-c_{\cal Q}\big[V(t)-\partial_t\phi(t)/2\big]\,.\label{eq:Density}
\end{align}

The apparent similarities between Eqs.~\eqref{eq:Current} and~\eqref{eq:Density} imply that the charge and current responses can be unified by here introducing the relativistic three-current given as $J_\mu=D_{\mu\nu}\big(A^\nu-\partial^\nu\phi/2\big)$ with $\mu,\nu=0,1,2$, metric tensor ${\rm diag}\{1,-1,-1\}${, and $D_{00}=-c_{\cal Q}$.} Hence, the superfluid stiffness and the JQC are proportional in systems with an emergent Lorentz invariance. Therefore, for Dirac electrons with a ``speed of light" $\upsilon_D$, Lorentz invariance leads to the constraint $D_{xx,yy}=\upsilon_D^2c_{\cal Q}$, which indeed holds for Eq.~\eqref{eq:CapacitanceUniversal}. Having settled the connection between $D$ and $c_{\cal Q}$ in SDSs, we now proceed by examining how nontrivial to\-po\-lo\-gy further imposes their nonuniversal quantization.

Our starting point is the Hamiltonian for a generic two-dimensional time-reversal invariant SC:
\begin{align}
\hat{H}(\bm{p})=\hat{h}(\bm{p})\tau_3+\Delta\tau_1,\label{eq:GenHam}
\end{align}

\noi where $\tau_{1,2,3}$ denote Pauli matrices which are defined in Nambu space. The latter is spanned by electrons with spin up and momentum $\bm{p}$, and, their time-reversed hole partners with spin down and momentum $-\bm{p}$. From Eqs.~\eqref{eq:Current} and~\eqref{eq:Density} we infer that $D_{\mu\nu}$ can be identified with the coefficients relating the uniform curent $J_{\mu}$ induced by a nonzero spatiotemporally-uniform phase gradient $\partial_\mu\phi$. {\color{black}In particular, to obtain expressions for the superfluid stiffness tensor elements $D_{ij}$, we consider small deviations of the superconducting phase away from the value $\phi=0$, i.e., $\phi(\bm{r})\approx(\partial_x\phi)x+(\partial_y\phi)y$, with $\partial_{x,y}\phi$ being constants. Subsequently, we phase twist the pai\-ring term according to $\Delta\tau_1\mapsto\Delta\tau_1e^{-i\phi(\bm{r})\tau_3}$ and determine the spatially-uniform current component $J_i$ within linear response to the phase gradient $\partial_j\phi$. In the same spirit, $D_{00}$ is found using linear response theory to a linearly-varying time-dependent phase of the form $\phi(t)\approx(\partial_t\phi)t$.  }

By adopting this alternative approach, in Ref.~\onlinecite{WJAextended} we show that $D_{\mu\nu}$ can be expressed as suitable response coefficients of the respective adiabatic Hamiltonian:
\begin{align}
\hat{\cal H}(\bm{p},\phi)=\hat{h}(\bm{p})\tau_3+\Delta\tau_1e^{-i\phi\tau_3}
\label{eq:AdiabaticHamiltonian}
\end{align}

\noi where the superconducting phase $\phi$ is now viewed as an additional synthetic momentum. Using the above adiabatic Hamiltonian, Ref.~\onlinecite{WJAextended} provides concrete expressions for the elements $D_{\mu\nu}$. {\color{black}
Specifically, for $D_{ij}$ we find~\cite{WJAextended}:
\begin{align}
D_{ij}=2\int dP\ph{\rm Tr}\Big[\hat{\upsilon}_i(\bm{p})\mathds{1}_\tau\hat{\cal F}_{p_j\phi}(\epsilon,\bm{p},\phi)\Big],\label{eq:Dform}
\end{align}

\noi where ``Tr'' denotes trace over all internal degrees of freedom. For compactness, we employed the shorthand notation
$\int dP\equiv\int_{\rm BZ}\frac{d\bm{p}}{(2\pi)^2}\int_{-\infty}^{+\infty}\frac{d\epsilon}{2\pi}$, with $\epsilon\in(-\infty,+\infty)$ corresponding to the frequency in imaginary time. The momenta are generally defined in a two-dimensional Brillouin zone (BZ), since the SC is considered to be a crystalline material. When eva\-lua\-ting the superfluid stiffness carried by a Dirac cone, we replace the BZ by $\mathbb{R}^2$~\cite{WJAextended}.

In Eq.~\eqref{eq:Dform}, we also introduced the group velocity $\hat{\bm{\upsilon}}(\bm{p})=\partial_{\bm{p}}\hat{h}(\bm{p})$ of the normal phase Hamiltonian $\hat{h}(\bm{p})$, along with the adiabatic matrix Green function:
\begin{align}
\hat{{\cal G}}^{-1}(\epsilon,\bm{p},\phi)=i\epsilon+B-\hat{{\cal H}}(\bm{p},\phi)\,.\label{eq:SynthGreen}
\end{align}

\noi We note once again that the Zeeman energy $B$ sets the Bogoliubov-Fermi level. The last ingredient of the adiabatic approach is the matrix Berry curvature function:
\begin{align}
\hat{\cal F}_{p_j\phi}=\nicefrac{1}{2}\big(\partial_\epsilon\hat{{\cal G}}^{-1}\big)\hat{{\cal G}}\big(\partial_\phi\hat{{\cal G}}^{-1}\big)\hat{{\cal G}}\big(\partial_{p_j}\hat{{\cal G}}^{-1}\big)\hat{{\cal G}}-\partial_\phi\leftrightarrow\partial_{p_j}\,,
\end{align}

\noi where we suppressed the arguments of the Green functions for notational convenience. In spite of the fact that our adiabatic approach is fully-equivalent to the standard method of evaluating the superfluid stiffness, it is unique in uncovering the possible underlying topological pro\-per\-ties of the system, since it is already expressed in the terms of a curvature function in a synthetic space~\cite{WJAextended}.}

We now proceed by recovering the result of Ref.~\onlinecite{KopninSoninPRL}  shown in Eq.~\eqref{eq:StiffnessUniversal}, and proving that $\pi D_{\rm cone}/\Delta$ is a topological invariant. At charge neutrality, the chemical potential $\mu$ is set to be zero and the normal phase Hamiltonian for a given graphene valley $\lambda=\pm1$ reads as~\cite{Graphene}:
\begin{align}
\hat{h}_\lambda(\bm{p})=\upsilon_D\big(p_x\sigma_1+\lambda p_y\sigma_2\big)
\label{eq:GrapheneNormal}
\end{align}

\noi where $\sigma_{1,2,3}$ denote Pauli matrices acting in the sublattice space spanned by the two interpenetrating triangular lattices of graphene~\cite{Graphene}. Note that, for graphene, electrons couple to holes of different valleys~\cite{Carlo}. Nonetheless, by a suitable choice of basis, each valley of graphene is described by a Hamiltonian of the form shown in Eq.~\eqref{eq:GenHam}.

The adiabatic Hamiltonian obtained for the $\lambda=+1$ valley of the graphene model in Eq.~\eqref{eq:GrapheneNormal} takes the form:
\begin{align}\hat{\cal H}_{\rm cone}(\bm{p},\phi)=\upsilon_D(p_x\sigma_1+p_y\sigma_2)\tau_3+\Delta\mathds{1}_\sigma\tau_1e^{-i\phi\tau_3}.
\end{align}

\noi Notably, the operator $\hat{\Pi}=\sigma_3\tau_3$ establishes a chiral symmetry $\big\{\hat{\cal H}_{\rm cone}(\bm{p},\phi),\hat{\Pi}\big\}=\hat{0}$. Its
presence guarantees that there exists a basis in which $\hat{\Pi}=\mathds{1}_\sigma\tau_3$ and the adiabatic Hamiltonian becomes block-off diagonal according to $\hat{\cal H}_{\rm cone}(\bm{p},\phi)=\hat{A}(\bm{p},\phi)\tau_++\hat{A}^{\dag}(\bm{p},\phi)\tau_-$~\cite{Schnyder2008}. Here, we introduced the off-diagonal matrices $\tau_\pm=(\tau_1\pm i\tau_2)/2$, along with the upper off-diagonal Hamiltonian block:
\begin{align}
\hat{A}(\bm{p},\phi)=\big(-\upsilon_Dp_x,-\upsilon_Dp_y,\Delta\cos\phi\big)\cdot\bm{\sigma}+i\Delta\sin\phi\mathds{1}_\sigma\,.
\end{align}

\noi The topological properties of the adiabatic Hamiltonian are encoded in the topological invariant $w_3\in\mathbb{Z}$~\cite{Schnyder2008}. This coincides with the winding number of $\hat{A}(\bm{p},\phi)$. We now write $w_3=\int_0^{2\pi}d\phi\ph w_3(\phi)/2\pi$, where we defined the win\-ding number densities $w_3(\phi)=\int d\bm{p}\ph w_3(\bm{p},\phi)/2\pi$ and:
\begin{align}
w_3(\bm{p},\phi)={\rm tr}\Big[\big(\hat{A}\partial_{p_x}\hat{A}^{-1}\big)\,\big(\hat{A}\partial_{p_y}\hat{A}^{-1}\big)\,\big(\hat{A}\partial_{\phi}\hat{A}^{-1}\big)\Big].
\label{eq:winding3densitymomentum}
\end{align}

\noi Notably, in the case of Dirac systems, $w_3(\bm{p},\phi)$ and $w_3(\phi)$ are independent of $\phi$. Consequently, besides $w_3$, also $w_3(\phi)$ is quantized. In particular, for a phase twist of $2\pi$ we have $w_3(\phi)=w_3$. The behaviors of $w_3(\bm{p},\phi)$ and $w_3(\phi)$ become relevant here since, for $B=0$, we find~\cite{WJAextended}:
\begin{align}
D_{\rm cone}=-\int\frac{d\bm{p}}{(2\pi)^2}\, E(\bm{p})\, w_3(\bm{p},\phi)\,,\label{eq:Dtopo}
\end{align}

\noi where $E(\bm{p})=\sqrt{(\upsilon_D\bm{p})^2+\Delta^2}$ and we made use of the fact that $D_{xx,yy}=D_{\rm cone}$ and $D_{xy,yx}=0$.

The above result highlights that the outcome for the superfluid stiffness is determined by the topological pro\-per\-ties of the adiabatic Dirac Hamiltonian. In fact, we are now in a position to prove that $\pi D_{\rm cone}/\Delta$ is a topological invariant itself. For this purpose, we note that for the evaluation of $w_3(\bm{p},\phi)$ we can linearize $\hat{A}$ with respect to $\phi$ according to $\hat{A}(\bm{p},\phi)\simeq\bm{g}(\bm{p})\cdot\bm{\sigma}+i\Delta\phi\mathds{1}_\sigma$. Here, we defined the vector $\bm{g}(\bm{p})=\big(-\upsilon_Dp_x,-\upsilon_Dp_y,\Delta\big)$, where $|\bm{g}(\bm{p})|=E(\bm{p})$. Under these assumptions, we find:
\begin{align}
\frac{D_{\rm cone}}{\Delta/\pi}=\int\frac{d\bm{p}}{4\pi i}E(\bm{p}){\rm tr}\left\{\hat{A}_0^{-1}(\bm{p})\big[\partial_{p_x}\hat{A}_0(\bm{p})\big]\big[\partial_{p_y}\hat{A}_0^{-1}(\bm{p})\big]\right\}
\end{align}

\noi where we took into account that $\phi$ can be set to zero after $\partial_\phi$ is carried out. In the above, we introduced $\hat{A}_0(\bm{p})\equiv\hat{A}(\bm{p},\phi=0)=\bm{g}(\bm{p})\cdot\bm{\sigma}$ which is a hermitian matrix with $\hat{A}_0^{-1}(\bm{p})=\hat{A}_0(\bm{p})/|\bm{g}(\bm{p})|^2$. The antisymmetry of the integrand under the exchange $p_x\leftrightarrow p_y$ further allows to obtain the expression:
\begin{align}
\frac{D_{\rm cone}}{\Delta/\pi}=2\int\frac{d\bm{p}}{4\pi}\ph\hat{\bm{g}}(\bm{p})\cdot\Big[\partial_{p_x}\hat{\bm{g}}(\bm{p})\times\partial_{p_y}\hat{\bm{g}}(\bm{p})\Big]\,,
\end{align}

\noi with the unit vector $\hat{\bm{g}}(\bm{p})=\bm{g}(\bm{p})/|\bm{g}(\bm{p})|$. When the momentum space is compact, the above integral yields an integer, that coincides with the 1st Chern number of the negative eigenstate of $\hat{A}_0(\bm{p})$. However, due to the Dirac nature of the adiabatic Hamiltonian, the integral yields $1/2$~\cite{VolovikBook}. Hence, the quantization of the superfluid stif\-fness directly probes the presence of a Weyl point at the origin of the synthetic coordinate space $(p_x,p_y,\Delta)$, since it is exactly there where $\hat{A}_0(\bm{p})$ and $|\bm{g}(\bm{p})|$ become zero.

The origin of the nontrivial topology can be traced back to the properties of the Hamiltonian in the absence of a phase bias: $\hat{H}_{\rm cone}(\bm{p})=\upsilon_D(p_x\sigma_1+p_y\sigma_2)\tau_3+\Delta\mathds{1}_\sigma\tau_1$. This Hamiltonian features two chiral symmetries effected by the operators $\hat{\Pi}=\sigma_3\tau_3$ and $\hat{\Pi}'=\tau_2$, which lead to a unitary symmetry with the generating ope\-ra\-tor $\hat{\cal O}=\sigma_3\tau_1$. {Employing} the unitary transformation $\big(\hat{\Pi}+\tau_1\big)/\sqrt{2}$ block dia\-go\-na\-li\-zes the unitary symmetry ope\-ra\-tor according to $\hat{\cal O}_\tau=\tau\mathds{1}_\sigma$ and the Hamiltonian into the blocks $\hat{H}_\tau(\bm{p})=\tau\bm{g}(\bm{p})\cdot\bm{\sigma}$. Since the occupied bands of the two massive Dirac Hamiltonians yield opposite fractional 1st Chern numbers with values $\tau/2$, we conclude that the superfluid stiffness is proportional to the dif\-fe\-ren\-ce of these two 1st Chern numbers. The latter can be viewed as a fractional ``spin"~\cite{Sheng} or dipole~\cite{WJAextended} 1st Chern number.

\begin{figure*}[t!]
\begin{center}
\includegraphics[width=1\textwidth]{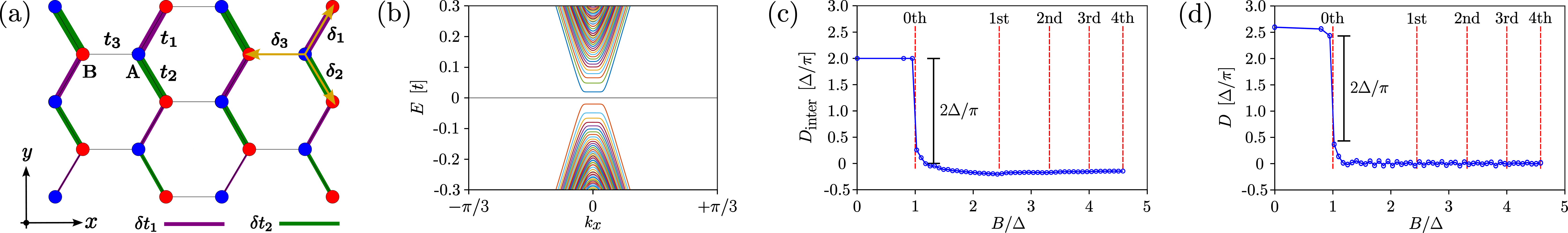}
\end{center}
\caption{(a) Schematic illustration of the model adopted for an armchair graphene nanoribbon (GNR). We consider nonuniform strain which renders the $x$ component of the pseudo-vector potential nonzero and increasing with increasing $y$, as indicated by the thickness of the green and purple lines. (b) Electronic band structure of (a) as a function of the conserved wave number $k_x$. First, we numerically obtain the spectrum for an armchair graphene GNR of width $W=601\times\sqrt{3}a_0$, where $a_0$ is the carbon-carbon distance, and $\omega_{\cal B}\simeq0.045$. Subsequently, we add a conventional superconducting gap $\Delta = 0.02$. pLLs emerge in the interval $-W/2\ell_{\cal B}^2\leq k_x\leq+W/2\ell_{\cal B}^2$ of the first Brillouin zone where $a_0$ is set to unity for convenience. (c) and (d) show numerical results for the interband and total superfluid stiffness components as functions of an inplane Zeeman field $B$ for the same values discussed in (b). The dashed vertical red lines indicate the energies for the pLLs in the presence of the nonzero $\Delta$. From Fig.~(c) we verify that $D_{\rm inter}\approx \frac{2\Delta}{\pi}$ for $B=0$. Across $B/\Delta=1$, the total superfluid stiffness drops by approximately $2\Delta/\pi$ as shown in (d). All energies are expressed in units of $t$, i.e., the nearest neighbor hopping in the absence of strain. {\color{black} Finally, for the numerics we have replaced the delta function entering in Eq.~\eqref{eq:Dintra} by a Lorentzian with width $\Gamma=0.001$.}}
\label{fig:Figure2}
\end{figure*}

It is important to examine the robustness of the quantized value of the superfluid stiffness against external perturbations. We first investigate situations which preserve chiral symmetry, as for instance the introduction of disorder in the pairing gap and the application of strain. {\color{black}The impact of weak and uncorrelated spatial disorder in the pairing gap} is analyzed in our supplementary file~\cite{SM} within the first-order Born approximation~\cite{PascalSimon}. We find that the quantization persists, but in the disordered case $\Delta$ is replaced by its spatially averaged value.

We now consider the presence of strain which varies li\-near\-ly in space. This scenario is particularly inte\-re\-sting, since the low-energy description of strained graphene solely consists of pLLs which possess a perfectly flat dispersion{\color{black}, therefore allowing us to make connections with topological constraints associated with quantum geo\-metry}. We adopt a strain profile which conserves the $p_y$ momentum and yields the Hamiltonian~\cite{Graphene,Salerno,Jamotte,SM}:
\bea
\hat{h}_\lambda^{\cal B}(p_y)=\omega_{\cal B}\left[\frac{\ell_{\cal B}}{\sqrt{2}}\hat{p}_x\sigma_1+\frac{1}{\sqrt{2}\ell_{\cal B}}\big(x+\lambda p_y\ell_{\cal B}^2\big)\sigma_2\right],
\label{eq:GrapheneStrain}
\eea

\noi where $\omega_{\cal B}=\sqrt{2}\upsilon_D/\ell_{\cal B}$. Here, $\ell_{\cal B}$ denotes the pseudomagnetic length. Each valley supports a single 0pLL which, for the given choice of strain profile, is an eigenstate of $\sigma_3$ with eigenvalue $-1$. The re\-mai\-ning spectrum of $\hat{h}_\lambda^{\cal B}(p_y)$ consists of two families of non-zero-energy pLLs, with eigenenergies $\varepsilon_{\sigma,n}(p_y)=\sigma\varepsilon_n(p_y)=\sigma\omega_{\cal B}\sqrt{n}$ for $n\geq1$. Each pLL sees a degeneracy per area which is equal to $1/2\pi\ell_{\cal B}^2$ for a single valley and a single spin projection~\cite{Graphene}.

{\color{black}Since the energy dispersions are flat, it is more convenient to deduce the contribution of each pLL to the stiffness using the standard expression for the interband part $D_{\rm inter}$~\cite{PeottaTormaBAB}. In the following we restrict to the $\lambda=+1$ valley and denote the corresponding pLL eigenvectors as $\left|u_\alpha(p_y)\right>$, where $\alpha=(\sigma,n)$ compactly labels the two quantum numbers. As we show in Ref.~\onlinecite{SM}, $D_{\rm inter}$ can be expressed as a sum of contributions defined per level $\alpha$:
\begin{align}
D_{\rm inter}^{\alpha}=\frac{\Delta}{\pi}\int\frac{dp_y}{W}\,\big<\partial_{p_y}u_\alpha(p_y)\big|\hat{M}_\alpha(p_y)\big|\partial_{p_y}u_\alpha(p_y)\big>P_\alpha(p_y)
\label{eq:Dgeom}
\end{align}

\noi where $W$ is the width of the sample. Here, $E_\alpha(p_y)=\sqrt{\varepsilon_\alpha^2(p_y)+\Delta^2}$, while $P_\alpha(p_y)= \Theta[E_\alpha(p_y)-|B|]$ controls the occupation of each level, where $\Theta$ is the Hea\-vi\-si\-de step function. The operator $\hat{M}_\alpha(p_y)$ is defined as follows:
\begin{align}
\hat{M}_\alpha(p_y)=\frac{2\Delta}{E_\alpha(p_y)}\frac{\hat{h}_{\lambda=+1}^{\cal B}(p_y)-\varepsilon_\alpha(p_y)}{\hat{h}_{\lambda=+1}^{\cal B}(p_y)+\varepsilon_\alpha(p_y)}\,.
\end{align}

\noi Equation~\eqref{eq:Dgeom} holds as long as $\hat{h}_\lambda^{\cal B}(p_y)+\varepsilon_\alpha(q)$ for all pLLs, which is exactly the case here as we explain in Ref.~\onlinecite{SM}.

By employing Eq.~\eqref{eq:Dgeom} at charge neutrality and $B=0$, we confirm that now only the 0pLLs contribute to the stiffness, which is still given by Eq.~\eqref{eq:StiffnessUniversal}. However, the stiffness is now purely geo\-me\-tric. To show this, we use that $\hat{M}_{{\rm 0pLL}}(p_y)=2\big(\hat{\mathds{1}}-\big|u_{\rm 0pLL}(p_y)\big>\big<u_{\rm 0pLL}(p_y)\big|\big)$. This allows us to express the stiffness in terms of the quantum metric of the 0pLL, i.e., $g_{\rm 0pLL}(p_y)=\big<\partial_{p_y}u_{\rm 0pLL}(p_y)\big|\big(\hat{\mathds{1}}-\big|u_{\rm 0pLL}(p_y)\big>\big<u_{\rm 0pLL}(p_y)\big|\big)\big|\partial_{p_y}u_{\rm 0pLL}(p_y)\big>=\ell_{\cal B}^2/2$. The result of Eq.~\eqref{eq:StiffnessUniversal} is recovered by evaluating the integral $\int dp_y/W$ by accounting for the degeneracy of the pLLs.
}

The arising robustness of the superfluid stiffness for weak strains is indeed expected  as long as the Dirac points in $(\bm{p},\phi)$ space of the adiabatic Hamiltonian persist. This holds when the pLL de\-ge\-ne\-ra\-cy is much smaller than unity, since then, the Dirac point is not removed but only ``moves'' in synthetic space. As a matter of fact, our adia\-ba\-tic approach is still valid for weak strains and we can re-evaluate Eq.~\eqref{eq:Dtopo} for $p_y\mapsto p_y+\lambda x/\ell_{\cal B}^2$ and $x\mapsto \phi/\partial_x\phi$. We find that corrections to Eq.~\eqref{eq:StiffnessUniversal} become negligible as long as $\omega_{\cal B}\ll\Delta$ and $\ell_{\cal B}$ is sufficiently smaller than the sample's  width $W$ for a strip geometry.

So far we have restricted to an ideal (un)strained Dirac cone. However, in realistic expe\-ri\-ments the nonconical part in graphene's band structure also contributes and spoils the universal behavior. Nonetheless, it is still possible to isolate the quantized part by introducing an inplane Zeeman field, which can selectively modify the occupation of the various states. In Fig.~\ref{fig:Figure2} we present numerical results for the superfluid stif\-fness of strained armchair graphene nanoribbons (GNRs) at charge neutrality and varying $B$. Notably, when $B$ crosses the ener\-gy bands, the arising Bogoliubov-Fermi points lead to an additional contribution to the intraband part of the superfluid stiffness{\color{black}, with the latter being obtained using the formula~\cite{SM}:
\begin{align}
D_{\rm intra}=-\frac{\Delta}{\pi}\int\frac{dp_y}{W}\sum_{\alpha}\frac{\Delta\big[\partial_{p_y}\varepsilon_\alpha(p_y)\big]^2}{2E_\alpha(p_y)}\frac{d}{dE_\alpha(p_y)}\frac{P_\alpha(p_y)}{E_\alpha(p_y)}.\label{eq:Dintra}
\end{align}
}

The results of Fig.~\ref{fig:Figure2} {\color{black}verify} that fields $|B|>\Delta$ ena\-ble the experimental detection of the quantized contribution of the 0pLLs. In Fig.~\ref{fig:Figure2}(d) we confirm the quantized jump of the total stiffness across $|B|=\Delta$, where any deviation from the expected quantization is only due to nu\-me\-ri\-cal finite-size effects. Our analysis reveals that, armchair GNRs are better-suited for observing the quantized jump compared to zigzag GNRs. This is because zigzag GNRs harbor additional edge flat bands which appear even without strain~\cite{Graphene} and spoil the quantization~\cite{SM}.

{\color{black}
Another aspect that remains to be addressed concerns the impact of detuning away from charge neutra\-li\-ty due to a nonzero chemical potential $\mu$ which leads to $\varepsilon_{\rm 0pLL}^{\mu\neq0}(p_y)=-\mu$ and $\varepsilon_{\sigma,n}^{\mu\neq0}(p_y)=\sigma\omega_{\cal B}\sqrt{n}-\mu$ for $n\geq1$ and $\sigma=\pm1$. Now, all pLLs contribute to the stiffness. Specifically, the contribution of the 0pLLs now becomes:
\begin{align}
D_{\lambda,0{\rm pLL}}^{\mu\neq0}=\frac{\Delta}{\pi}\frac{1}{\sqrt{1+\big(\mu/\Delta\big)^2}}\frac{1}{1-\left(\frac{\mu}{\omega_{\cal B}/2}\right)^2}\,.
\end{align}

\noi Besides a renormalized $\Delta$, an additional factor emerges which diverges for $|\mu|=\omega_{\cal B}/2$. Thus, the stiffness can be strongly enhanced by tuning the system to this resonance. In fact, such re\-so\-nan\-ces appear for all pLLs~\cite{SM}.
}

{\color{black}After investigating the superfluid stiffness,} we now evaluate the quantum capacitance~\cite{Brey,Ouyang,Jena,GrapheneCQ,Ensslin,Ponomarenko} for a Josephson junction fabricated by con\-tac\-ting {\color{black}an ideal} strained monolayer graphene hybrid~\cite{Morpurgo,Andrei,Choi,Mizuno,Calado,Finkelstein,Shalom,Bretheau,Seredinski} to a conventional SC, as shown in Fig.~\ref{fig:Figure1}(b). Using the model in Eq.~\eqref{eq:GrapheneStrain} for strained graphene{\color{black}, we find that the two-valley JQC at charge neutrality and $B=0$ takes the exact form~\cite{SM}:}
\begin{align}
c_{\cal Q}^{B=\mu=0}= 
2\cdot\frac{\Delta}{\pi\upsilon_D^2}\left\{\left(\frac{\Delta}{2\omega_{\cal B}}\right)\zeta\left[\nicefrac{3}{2},\left(\Delta/\omega_{\cal B}\right)^2\right]-\left(\frac{\omega_{\cal B}}{2\Delta}\right)^2\right\}
\end{align}

\noi with the Hurwitz zeta function $\zeta$. Thus, at charge neutra\-li\-ty, weak strains $\omega_{\cal B}\ll\Delta$ yield $c_{\cal Q}\rightarrow2\cdot\Delta/(\pi\upsilon_D^2)\equiv c_0$.  {\color{black}This result marks the topological regime and is in accordance with the value set by the emergent Lorentz invariance. In contrast}, for $\omega_{\cal B}\gg\Delta$ only the 0pLLs contribute with $c_{\cal Q}/c_0\rightarrow (\omega_{\cal B}/2\Delta)^2\gg1$. For $\Delta=0.2\ph{\rm meV}$ and $\upsilon_D=10^6\ph{\rm m/s}$, we find $c_0\simeq5\ph{\rm nF/cm^2}$. Therefore, tai\-loring the classical ca\-pa\-ci\-tance of the junction $c_{\PK{C\ell}}$, so that $c_{\PK{C\ell}}\gg c_{\cal Q}$, in principle enables the detection of the underlying properties of the Dirac points. This may be possible by obser\-ving Coulomb-bloc\-kade-induced char\-ging effects, which can be controlled by means of strain~\cite{SM} and Zeeman field~\cite{WJAextended} engineering.

We conclude by discussing broader implications of the topological quantization of $D$ and $c_{\cal Q}$ demonstrated here. First of all, the invariance of the two response coefficients for weak strains implies that flat band SCs which are dictated by a solely-geometric quantized stiffness, can be adiabatically {\color{black}mapped to SDSs and the arising quantization linked to the quantum metric can be understood via topological constraints on the total SDS superfluid stiffness. Even more}, based on the observation that moir\'e twisting can be effectively viewed as the application of strain~\cite{XiDai}, we infer that the nonstandard topological properties proposed {\color{black}here for strained SDSs can be also applicable to their moir\'e-twisted counterparts.

In this spirit, we expect that} the geo\-me\-tric superfluid stiffness of MATBG~\cite{Ahn,Randeria,Pikulin,Julku,BABbounds,BABbounds2} {\color{black}can be possibly} attributed to the topological superfluid stiffness of  untwisted bilayer graphene. {\color{black}Although we leave the verification of the above conjecture for a future work~\cite{MATBGcomment}, we here stress that such a scenario is indeed} plausible because the total superfluid stiffness of a number of $s$ uncoupled graphene layers, or more ge\-ne\-ral of a number of $s$ uncoupled Dirac cones, sa\-ti\-sfies the quantization law $D^{(s)}=|s|D_{\rm cone}$~\cite{WJAextended}. Therefore, the here-proposed connection between Dirac cones and flat band systems opens the door to predicting and linking {\color{black}distinct} quantum materials and devices which {\color{black}yet share the same} topological superfluid responses.

Aside from the above concept, the nonuniversal topological quantization for $D^{(s)}$ further implies the \textit{universal} quantization of the admittance modulus $Y_{\rm mod}$ of the SDS for $\omega=2\Delta$, i.e., at the absorption edge. This relies on the fact that for ${\color{black}0<}\omega\leq2\Delta$ and at zero temperature, a fully-gapped nondisordered SC behaves as a perfect inductor with $Y_{\rm mod}=D/\omega$~\cite{Tinkham}. {\color{black}Thus}, in our case we find:
\begin{align}
Y_{\rm mod}^{(s)}\big(\mu=0,{\color{black}0<}\omega\leq2\Delta\big)=|s|\frac{e^2}{h}\frac{2\Delta}{\hbar \omega}\,,
\label{eq:QAE}\end{align}

\noi where the quantum of conductance $e^2/h$ appeared after restoring $e$ and $\hbar$. From the above, we confirm that a quantum admittance effect (QAE) emerges for $\hbar\omega=2\Delta$.

Currently, however, it is challenging to experimentally observe the above QAE in graphene-SC hybrids, since the chemical potential can be only reduced down to $\sim1\,{\rm meV}$~\cite{Mayorov}. This sets exi\-sting platforms in the antipodal regime $|\mu|\gg\Delta$, thus, raising the question of how does the stiffness and the admittance behave in this limit. Re-evaluating the  stiffness for $\mu\neq0$ yields that a higher-order Dirac cone with vorticity  $s$ carries a stiffness $D^{(s)}(|\mu|\gg\Delta)\approx|s||\mu|/2\pi$~\cite{SM}. In fact, for such a higher-order SDS, the stiffness is proportional to $|s|$ for all $\mu$~\cite{WJAextended}. Since a gate potential $V_g$ modifies $\mu$ according to $\mu+V_g$, we propose to experimentally measure $dD^{(s)}/dV_g\approx |s|/2\pi$ which, while it is not a topological invariant, it is still approximately universally  quantized.

\section*{Acknowledgements}

We are thankful to Mario Cuoco, Maria Teresa Mercaldo, Tao Shi, and Hong-Qi Xu for helpful discussions. We acknowledge funding from the National Na\-tu\-ral Scien\-ce Foundation of China (Grant No.~12074392).

\newpage

\appendix

\begin{widetext}

\section{{\large Supplemental Material}}
\section{ Topological Superfluid Responses of Superconducting Dirac Semimetals}

\subsection{Jun-Ang Wang$^{1,2}$, Mohamed Assili$^1$, and Panagiotis Kotetes$^1$}
\begin{center}
$^1$ CAS Key Laboratory of Theoretical Physics, Institute of Theoretical Physics,\\
Chinese Academy of Sciences, Beijing 100190, China\\
$^2$ School of Physical Sciences, University of Chinese Academy of Sciences, Beijing 100049, China
\end{center}

\section{{A. Effects of Disorder on the Quantization of the Superfluid Stiffness}}

It is important to investigate the robustness of the quantization of the superfluid stiffness for a single Dirac cone in the presence of disorder in the pairing gap. This is crucial, since the pairing gap sets the superfluid stiffness quantum. For this purpose, we consider that the pairing gap becomes modified as $\Delta\mapsto\Delta[1+f(\bm{r})]$, where the function $f(\bm{r})$ is considered to give rise to spatially uncorrelated disorder and can be equivalently expressed using the Fourier decomposition $f(\bm{r})=\int\frac{d\bm{q}}{(2\pi)^2}\, e^{i\bm{q}\cdot\bm{r}}f(\bm{q})$. In the remainder, we focus on weak disorder strengths by restricting to $|f(\bm{r})|\ll1$. Under this assumption, it is eligible to employ the Born approximation. We thus infer the self-energy $\hat{\Sigma}(\epsilon,\bm{p})$ of the single particle Green function $\hat{G}(\epsilon,\bm{p})$ due to disorder. This Green function is defined for the Hamiltonian in Eq.~(5) of the main text through the formula $\hat{G}^{-1}(\epsilon,\bm{p})=i\epsilon+B-\hat{H}(\bm{p})$. Note that here it is also eligible to restrict ourselves to the study of the time-ordered Green function, instead of the retarded one, because we assume that the disorder strength is sufficiently weak to guarantee that possible level broadenings are much smaller than the pairing gap $\Delta$. The validity of such an approach in the present case can be inferred from the analysis of prior studies on Dirac superconductors, e.g., see Ref.~\onlinecite{PascalSimon}. According to this work, the Born approximation without selfconsistency is sufficient for examining the effects of disorder in this regime. Hence, under this condition and within the nonselfconsistent Born approximation (BA), we find the expression for the self-energy $\hat{\Sigma}_{\rm BA}(\epsilon,\bm{p})$ in the absence of a Zeeman field ($B=0$):
\begin{align}
\hat{\Sigma}_{\rm BA}(\epsilon,\bm{p})=\frac{\Delta^2}{(2\pi)^2}\int\frac{d\bm{q}}{(2\pi)^2}\,|f(\bm{q}-\bm{p})|^2\,\tau_1\hat{G}(\epsilon,\bm{q})\tau_1=-\frac{\Delta^2}{(2\pi)^2}\int\frac{d\bm{q}}{(2\pi)^2}\,|f(\bm{q}-\bm{p})|^2\,\frac{i\epsilon-\upsilon_D(q_x\sigma_1+q_y\sigma_2)\tau_3+\Delta\tau_1}{(\upsilon_D \bm{q})^2+\Delta^2+\epsilon^2}\,.
\end{align}

\noi Since in the present work we are mainly interested in the effects of disorder near the Dirac point at $\bm{p}=\bm{0}$, we can set $\hat{\Sigma}_{\rm BA}(\epsilon,\bm{p})\approx\hat{\Sigma}_{\rm BA}(\epsilon,\bm{p}=\bm{0})$. By further assuming slowly varying disorder, we have $f(\bm{q})\approx f(\bm{0})$. These assumptions allow us to make progress with analytical manipulations. We find that terms $\propto\tau_3$ do not survive the angular integration in momentum space and we thus end up with the expression:
\begin{align}
\hat{\Sigma}_{\rm BA}(\epsilon,\bm{p})\approx-\frac{|f(\bm{0})|^2}{(2\pi)^3}\left(\frac{\Delta}{\upsilon_D}\right)^2\int_0^\Lambda\,du\,u\,\frac{i\epsilon+\Delta\tau_1}{u^2+\Delta^2+\epsilon^2}=-\frac{\big[|f(\bm{0})|/\xi_{\rm sc}\big]^2}{(2\pi)^3}\ln\left(\frac{\sqrt{\Lambda^2+\Delta^2+\epsilon^2}}{\sqrt{\Delta^2+\epsilon^2}}\right)\big(i\epsilon+\Delta\tau_1\big)\,,
\end{align}

\noi where we introduced the superconducting coherence length $\xi_{\rm sc}=\upsilon_D/\Delta$ and an energy cutoff $\Lambda$. Since any topological properties are stemming from the vicinity of $\epsilon\sim0$, we can approximate $\sqrt{\Delta^2+\epsilon^2}\approx\Delta$. In addition, in the remainder we consider the case $\Lambda\gg\Delta$ which further allows us to proceed with the approximation $\sqrt{\Lambda^2+\Delta^2+\epsilon^2}\approx\Lambda$. By defining the disorder strength $\gamma=|f(\bm{0})|^2\ln\big(\Lambda/\Delta\big)/\big[\xi_{\rm sc}^2(2\pi)^3\big]\geq0$ we end up with the formula $\hat{\Sigma}_{\rm BA}(\epsilon,\bm{p})\approx-\gamma\big(i\epsilon+\Delta\tau_1\big)$ and the Green function becomes:
\begin{align}
\hat{G}_{\rm BA}^{-1}(\epsilon,\bm{p})=i\epsilon-\upsilon_D(p_x\sigma_1+p_y\sigma_2)\tau_3-\Delta\tau_1-\hat{\Sigma}_{\rm BA}(\epsilon,\bm{p})=Z^{-1}\big[i\epsilon-\tilde{\upsilon}_D(p_x\sigma_1+p_y\sigma_2)-\tilde{\Delta}\tau_1\big]\,,
\end{align}

\noi where we introduced the renormalization factor $Z^{-1}=1+\gamma$, along with the renormalized quantities $\tilde{\upsilon}_D=Z\upsilon_D$ and $\tilde{\Delta}=(2Z-1)\Delta$. Note that the $\tilde{\Delta}$ corresponds to the value of the pairing gap that one identifies using spectroscopic methods. In the situations of interest, $\tilde{\Delta}$ can be inferred by means of spatially averaging the observed gap edge in the density of states of superconducting graphene using scanning tunneling microscopy.

We now proceed with examining the impact of this type of spatial disorder on the superfluid stiffness. Firstly, we note that the paramagnetic current vertex, which is given by the term $\partial_{\bm{p}}\hat{G}_{\rm BA}^{-1}(\epsilon,\bm{p})\tau_3$,\footnote{In the most general case one has to restrict the momentum derivative to the nonsuperconducting part of the Green function. However, in the case of a momentum-independent pairing gap the expression for the vertex can be extended to include the full Green function.} remains identical to the one for the clean case. Therefore, the paramagnetic current operator in the disordered case $\hat{\bm{J}}_{\rm BA}^{(p)}(\bm{p})$ continues to be given by $\hat{\bm{J}}_{\rm BA}^{(p)}(\bm{p})=\hat{\bm{J}}_{\rm clean}^{(p)}(\bm{p})\equiv-\hat{\bm{\upsilon}}(\bm{p})$ even in the presence of the pairing gap disorder. This additionally implies that the superconducting phase simply modifies the pairing gap in this disordered case by means of the shift $\tilde{\Delta}\mapsto\tilde{\Delta}e^{i\phi(\bm{r})\tau_3}$. We are now in a position to define the adiabatic Green function in the absence of a Zeeman field according to $\hat{\cal G}_{\rm BA}^{-1}(\epsilon,\bm{p},\phi)=Z^{-1}\big[i\epsilon-Z\hat{h}(\bm{p})\tau_3-\tilde{\Delta}\tau_1e^{-i\phi\tau_3}\big]$. Next in line is to infer the modification on $\hat{\cal F}_{p_i,\phi}(\epsilon,\bm{p},\phi)$ in the presence of this kind of disorder. Following once again {\color{black}the same steps as in Ref.~\onlinecite{WJAextended}}, yields that $\hat{\cal F}_{{\rm BA};p_i,\phi}(\epsilon,\bm{p},\phi)=Z\hat{\cal F}_{{\rm clean};p_i,\phi}(\epsilon,\bm{p},\phi)$. This analysis implies that the superfluid stiffness can be now expressed as:
\begin{align}
D_{ij}^{\rm disorder}=2\int dP\ph{\rm Tr}\Big[\hat{\tilde{\upsilon}}_i(\bm{p})\mathds{1}_\tau\hat{\widetilde{{\cal F}}}_{p_j\phi}(\epsilon,\bm{p},\phi)\Big]\,,
\label{eq:StartingPointDis}
\end{align}

\noi with $\hat{\tilde{\bm{\upsilon}}}(\bm{p})=\partial_{\bm{p}}\hat{\tilde{h}}(\bm{p})$, where we introduced $\hat{\tilde{h}}(\bm{p})=Z\hat{h}(\bm{p})$, and $\hat{\widetilde{{\cal F}}}_{p_j\phi}(\epsilon,\bm{p},\phi)$ {\color{black}is obtained by employing} the adiabatic Green function $\hat{\widetilde{{\cal G}}}(\epsilon,\bm{p},\phi)=\big[i\epsilon-\hat{\tilde{h}}(\bm{p})\tau_3-\tilde{\Delta}\tau_1e^{-i\phi\tau_3}\big]^{-1}$. Notably, the latter does not include the renormalization factor $Z$ and is expressed solely in terms of the renormalized Hamiltonian quantities. The reason why we can obtain these simplified relations is because $\hat{\cal F}_{{\rm BA};p_i,\phi}(\epsilon,\bm{p},\phi)$ contains an equal number of Green functions and their inverses, so that the renormalization factors stemming from the Green functions cancel out. The remaining $Z$ factor appearing in $\hat{\cal F}_{{\rm BA};p_i,\phi}(\epsilon,\bm{p},\phi)$ is then absorbed by redefining the current vertex.

In conclusion, we find that in the presence of the disorder in the pairing gap, the superfluid stiffness is given by the expressions obtained for the clean case but with renormalized {\color{black}Dirac velocity} and pairing gap. Therefore, in the case of superconducting graphene, disorder in the pairing gap leads to a modified stiffness per Dirac cone which now becomes $\tilde{\Delta}/\pi$. Hence, the quantization persists, but the pairing gap is given by the disorder averaged pairing gap. As we mentioned earlier, $\tilde{\Delta}$ is actually the gap that one obtains experimentally using spectroscopic methods. We thus conclude that the clean and disordered cases do not essentially differ, as long as one defines the quantum of superfluid stiffness in terms of the experimentally observed pairing gap.

\section{B. Strained Graphene Hybrids - Analytics}

\noi In the case of strained graphene at neutrality (i.e. $\mu=0$) we have contributions from two valleys ($\lambda=\pm1$) and the respective Hamiltonians read as:
\bea
\hat{h}_\lambda(q)=\upsilon_D\left[\hat{p}_\perp\sigma_1+\big(q+\lambda x_\perp/\ell_{\cal B}^2\big)\lambda\sigma_2\right]=\omega_{\cal B}\left[\frac{\ell_{\cal B}}{\sqrt{2}}\hat{p}_\perp\sigma_1+\frac{1}{\sqrt{2}\ell_{\cal B}}\big(x_\perp+\lambda q\ell_{\cal B}^2\big)\sigma_2\right]\,,\no
\eea

\noi where $\omega_{\cal B}=\sqrt{2}\upsilon_D/\ell_{\cal B}$. We remark that in the above Hamiltonian and in the remainder of this file, we have made for notational convenience the replacements $x\mapsto x_\perp$ and $(p_x,p_y)\mapsto(p_\perp,q)$. We see that the Hamiltonian of one valley maps to the Hamiltonian of the other by taking $q\mapsto-q$. Therefore, in the following, we restrict for convenience to the valley with $\lambda=1$ and multiply by a factor of $2$ to account for both valleys in the final results.

\subsection{Eigenstates, Eigenenergies, and Matrix Elements}

We introduce the ladder operators for valley $\lambda=1$:
\bea
\hat{a}(q)=\frac{\ell_{\cal B}}{\sqrt{2}}\hat{p}_\perp-i\frac{x_\perp+q\ell_{\cal B}^2}{\sqrt{2}\ell_{\cal B}}\quad{\rm and}\quad
\hat{a}^\dag(q)=\frac{\ell_{\cal B}}{\sqrt{2}}\hat{p}_\perp+i\frac{x_\perp+q\ell_{\cal B}^2}{\sqrt{2}\ell_{\cal B}}
\no\eea

\noi which satisfy $[\hat{a}(q),\hat{a}^\dag(q)]=1$, and the Hamiltonian can be written as $\hat{h}(q)=\omega_{\cal B}\big[\hat{a}^\dag(q)\sigma_-+\hat{a}(q)\sigma_+\big]$, where $\sigma_\pm=(\sigma_1\pm i\sigma_2)/2$. We find that $\hat{h}^2(q)/\omega_{\cal B}^2=[\hat{a}^\dag(q)\sigma_-+\hat{a}(q)\sigma_+]^2=\hat{a}^\dag(q)\hat{a}(q)(1-\sigma_3)/2+\hat{a}(q)\hat{a}^\dag(q)(1+\sigma_3)/2=\hat{a}^\dag(q)\hat{a}(q)+(1+\sigma_3)/2$. With the given conventions $\hat{h}(q)$ observes a zero-energy Landau level for $\sigma_3=-1$ and eigenvector:
\bea
\big|u_{0,-1}(q)\big>=\big|\phi_0(q)\big>\left(\begin{array}{c}0\\1\end{array}\right)\quad{\rm with}\quad \varepsilon_{0,-1}=0\,,\no
\eea

\noi where we introduced the eigenstates of the displaced quantum harmonic oscillator $\big|\phi_n(q)\big>$ (with displacement $q\ell_{\cal B}^2$) which satisfy the defining relation $\hat{a}^\dag(q)\hat{a}(q)\big|\phi_n(q)\big>=n\big|\phi_n(q)\big>$.

Knowing the zero-energy Landau level allows us to determine the remaining spectrum of $\hat{h}(q)$, which is given in term of the following two families of non-zero-energy Landau levels:
\bea
\big|u_{n,\sigma}(q)\big>=\frac{1}{\sqrt{2}}\left(\begin{array}{c}\big|\phi_{n-1}(q)\big>\\\\\sigma\big|\phi_n(q)\big>\end{array}\right)\quad{\rm with}\quad \varepsilon_{n,\sigma}(q)=\sigma\varepsilon_n(q)=\sigma\omega_{\cal B}\sqrt{n}\quad{\rm for}\quad
n\geq1\,.\no
\eea

\noi Each Landau level sees a degeneracy $1/2\pi\ell_{\cal B}^2$ (per given valley and a single spin projection). The above set the stage for the evaluation of the current response. For this purpose it is required to evaluate the matrix elements $\big<u_{n,\sigma}(q)\big|\partial_qu_{n',\sigma'}(q)\big>$. Since the displaced quantum harmonic oscillator basis is the most convenient, we write $\partial_q\equiv\ell_{\cal B}^2\partial_x=\frac{i\ell_{\cal B}}{\sqrt{2}}\big(\sqrt{2}\ell_{\cal B}\hat{p}_\perp\big)=\frac{i\ell_{\cal B}}{\sqrt{2}}\big[\hat{a}(q)+\hat{a}^\dag(q)\big]$. The above implies for $n\geq1$ and $n'\geq1$ that:
\bea
&&-i\big(\sqrt{2}/\ell_{\cal B}\big)\big<u_{n,\sigma}(q)\big|\partial_qu_{n',\sigma'}(q)\big>=
\big<u_{n,\sigma}(q)\big|\big[\hat{a}(q)+\hat{a}^\dag(q)\big]\big|u_{n',\sigma'}(q)\big>\no\\\no\\
&&=\frac{\big<\phi_{n-1}(q)\big|\big[\hat{a}(q)+\hat{a}^\dag(q)\big]\big|\phi_{n'-1}(q)\big>+\sigma\sigma'
\big<\phi_n(q)\big|\big[\hat{a}(q)+\hat{a}^\dag(q)\big]\big|\phi_{n'}(q)\big>}{2}\no\\
&&=\frac{\sqrt{n}\delta_{n',n+1}+\sqrt{n'}\delta_{n'+1,n}+\sigma\sigma'\big(\sqrt{n'}\delta_{n',n+1}+\sqrt{n}\delta_{n'+1,n}\big)}{2}\no\\
&&=\big(\sigma\sqrt{n}+\sigma'\sqrt{n'}\big)\frac{\sigma\delta_{n',n+1}+
\sigma'\delta_{n'+1,n}}{2}
=\frac{\varepsilon_{n,\sigma}+\varepsilon_{n',\sigma'}}{\omega_{\cal B}}\frac{\sigma\delta_{n+1,n'}+
\sigma'\delta_{n'+1,n}}{2}
\,,\no
\eea

\noi while for $(n,\sigma)=(0,-1)$ and $n'\geq1$ we have:
\bea
-i\big(\sqrt{2}/\ell_{\cal B}\big)\big<u_{0,-1}(q)\big|\partial_qu_{n',\sigma'}(q)\big>=
\big<u_{0,-1}(q)\big|\big[\hat{a}(q)+\hat{a}^\dag(q)\big]\big|u_{n',\sigma'}(q)\big>
=\sigma'\frac{\big<\phi_0(q)\big|\big[\hat{a}(q)+\hat{a}^\dag(q)\big]\big|\phi_{n'}(q)\big>}{\sqrt{2}}=\frac{\sigma'\delta_{n',1}}{\sqrt{2}}\,.\no
\eea

\section{C. Superfluid Stiffness: Definitions and Calculations}

\noi This section relies on the analysis of our accompanying work in Ref.~\onlinecite{WJAextended}, after considering the quasi-one-dimensional case, for a system in a strip geometry with width $W$.

\subsection{Band-Defined Contributions}

We now introduce the eigenstates of $\hat{h}(q)$, which we label as $\big|u_\alpha(q)\big>$ with energy dispersions $\varepsilon_\alpha(q)$. We also set $E_\alpha(q)=\sqrt{\varepsilon_\alpha^2(q)+\Delta^2}$. The expression that leads to the superfluid stiffness reads as:
\bea
D=2\Delta^2\int_{-\pi}^{+\pi}\frac{dq}{2\pi W}\sum_{\alpha,\beta}\frac{\big|\big<u_\alpha(q)\big|\partial_q\hat{h}(q)\big|u_\beta(q)\big>\big|^2}{E_\alpha^2(q)-E_\beta^2(q)}\left[\frac{P_\beta(q)}{E_\beta(q)}-\frac{P_\alpha(q)}{E_\alpha(q)}\right]\,,\no
\eea

\noi where $P_\alpha(q)=\Theta[B+E_\alpha(q)]-\Theta[B-E_\alpha(q)]\equiv\Theta[E_\alpha(q)-|B|]$, and $\Theta(\epsilon)$ denotes the Heaviside step function. We now separate intra- and inter-band contributions. For $\alpha=\beta$ the matrix element $\big<u_\alpha(q)\big|\partial_q\hat{h}(q)\big|u_\beta(q)\big>$ is simply given by $\partial_q\varepsilon_\alpha(q)$. On the other hand, for $\alpha\neq\beta$ we exploit the relation $\big<u_\alpha(q)\big|\partial_q\hat{h}(q)\big|u_\beta(q)\big>=[\varepsilon_\beta(q)-\varepsilon_\alpha(q)]\big<u_\alpha(q)\big|\partial_qu_\beta(q)\big>$. The above considerations lead to the following expression for the intraband (conventional) contribution:
\bea
D_{\rm intra}=\frac{\Delta}{\pi}\int_{-\pi}^{+\pi}\frac{dq}{W}\sum_{\alpha}\frac{\Delta}{E_\alpha(q)}\frac{1}{2}\left[\frac{\partial_q\varepsilon_\alpha(q)}{E_\alpha(q)}\right]^2\Big\{P_\alpha(q)-E_\alpha(q)\delta\big[E_\alpha(q)-|B|\big]\Big\},\label{eq:Dintra}
\eea

\noi as well as to the relation for the interband one:
\bea
D_{\rm inter}=\frac{\Delta^2}{\pi}\int_{-\pi}^{+\pi}\frac{dq}{W}\sum_{\alpha,\beta}\big|\big<u_\alpha(q)\big|\partial_qu_\beta(q)\big>\big|^2\frac{\big[\varepsilon_\alpha(q)-\varepsilon_\beta(q)\big]^2}{E_\alpha^2(q)-E_\beta^2(q)}\left[\frac{P_\beta(q)}{E_\beta(q)}-\frac{P_\alpha(q)}{E_\alpha(q)}\right]\,.\no
\eea

\noi In addition, for systems in which the matrix elements between states with $\varepsilon_\beta(q)+\varepsilon_\alpha(q)=0$ are identically zero, we employ the relation $E_\alpha^2(q)-E_\beta^2(q)=\varepsilon_\alpha^2(q)-\varepsilon_\beta^2(q)$, and re-express the interband part as:
\begin{align}
D_{\rm inter}=\frac{\Delta}{\pi}\int_{-\pi}^{+\pi}\frac{dq}{W}\sum_{\alpha,\beta}2\big|\big<u_\alpha(q)\big|\partial_qu_\beta(q)\big>\big|^2\frac{\Delta}{E_\alpha(q)}\frac{\varepsilon_\beta(q)-\varepsilon_\alpha(q)}{\varepsilon_\beta(q)+\varepsilon_\alpha(q)}P_\alpha(q)\,,\no
\end{align}

\noi which can be compactly written as:
\bea
D_{\rm inter}=\frac{\Delta}{\pi}\sum_\alpha\int_{-\pi}^{+\pi}\frac{dq}{W}\ph\big<\partial_qu_\alpha(q)\big|\frac{2\Delta}{E_\alpha(q)}\frac{\hat{h}(q)-\varepsilon_\alpha(q)}{\hat{h}(q)+\varepsilon_\alpha(q)}\big|\partial_qu_\alpha(q)\big>P_\alpha(q)\,.\no
\eea

\noi Finally, we note that the contribution of a zero-energy flat band defined in the entire momentum space reads as:
\bea
D_0&=&\frac{\Delta}{\pi}P_0\int_{-\pi}^{+\pi}\frac{dq}{W}\ph2g_0(q)\,,\no\label{eq:StiffnessQM}
\eea

\noi where $P_0=\Theta(\Delta-|B|)$ and $g_0(q)=\big<\partial_qu_0(q)\big|\big[\hat{\mathds{1}}-\hat{{\cal P}}_0(q)\big]\big|\partial_qu_0(q)\big>$ defines the quantum metric tensor element of the flat band and $\hat{{\cal P}}_0(q)$ the respective projector onto it.

\subsection{Superfluid Stiffness at Charge Neutrality}

Having obtained the general expression for $D$ and the matrix elements between states of strained graphene within the Dirac-cone approximation, we proceed with the evaluation of the superfluid stiffness. Given that $\mu=0$ and by further considering $B=0$, we have the following:
\bea
D_{\rm inter}&=&
2\cdot 4\Delta\frac{\ell_{\cal B}^2}{2}\int_{-q_c}^{+q_c}\frac{dq}{2\pi W}\sum_{(n,\sigma)}\frac{\Delta}{E_{n,\sigma}}\sum_{(n',\sigma')}\frac{2}{\ell_{\cal B}^2}\big|\big<u_{n,\sigma}(q)\big|\partial_qu_{n',\sigma'}(q)\big>\big|^2\frac{\varepsilon_{n',\sigma'}-\varepsilon_{n,\sigma}}{\varepsilon_{n',\sigma'}+\varepsilon_{n,\sigma}}\no\\
&=&2\cdot\frac{\Delta}{\pi}\sum_{(n,\sigma)}\frac{\Delta}{E_{n,\sigma}}\sum_{(n',\sigma')}\big|\big<u_{n,\sigma}\big|\left(\hat{a}+\hat{a}^\dag\right)\big|u_{n',\sigma'}\big>\big|^2\frac{\varepsilon_{n',\sigma'}-\varepsilon_{n,\sigma}}{\varepsilon_{n',\sigma'}+\varepsilon_{n,\sigma}}\no
\eea

\noi where we accounted for both valleys by multiplying by a factor of $2$ and suitably replaced the integral over $q$ by means of considering the degeneracy of each Landau level. The energy scales appearing above read $\varepsilon_{n,\sigma}=\sigma\varepsilon_n=\sigma\omega_{\cal B}\sqrt{n}$ and $E_{n,\sigma}=E_n=\sqrt{\omega_{\cal B}^2n+\Delta^2}$.  Using the above expression, we can evaluate the contribution of each band to the superfluid stiffness. We first evaluate the contribution of the two zeroth pseudo-Landau levels:
\bea
D_{n=0,\sigma=-1}
=2\cdot\frac{\Delta}{\pi}\sum_{\sigma}\big|\big<u_{0,-1}\big|\left(\hat{a}+\hat{a}^\dag\right)\big|u_{1,\sigma}\big>\big|^2=
2\cdot\frac{\Delta}{\pi}\,.\no
\eea

\noi We continue with $n=1$:
\bea
D_{n=1,\sigma}
&=&2\cdot\frac{\Delta}{\pi}\frac{\Delta}{E_1}\left\{-\big|\big<u_{1,\sigma}\big|\left(\hat{a}+\hat{a}^\dag\right)\big|u_{0,-1}\big>\big|^2
+\sum_{\sigma'}\big|\big<u_{1,\sigma}\big|\left(\hat{a}+\hat{a}^\dag\right)\big|u_{2,\sigma'}\big>\big|^2\frac{\sigma'\sqrt{2}-\sigma}{\sigma'\sqrt{2}+\sigma}\right\}=0\,.\no
\eea

\noi We now complete the calculation with the evaluation of $n\geq2$.
\bea
D_{n\geq2,\sigma}
&=&2\cdot\frac{\Delta}{\pi}\frac{\Delta}{E_n}\sum_{(n',\sigma')}\left(\frac{\varepsilon_{n,\sigma}+\varepsilon_{n',\sigma'}}{\omega_{\cal B}}\right)^2\left(\frac{\sigma\delta_{n+1,n'}+
\sigma'\delta_{n'+1,n}}{2}\right)^2\frac{\varepsilon_{n',\sigma'}-\varepsilon_{n,\sigma}}{\varepsilon_{n',\sigma'}+\varepsilon_{n,\sigma}}\no\\
&=&2\cdot\frac{\Delta}{\pi}\frac{\Delta}{E_n}\sum_{n'}\frac{\delta_{n+1,n'}+
\delta_{n'+1,n}}{2}\big(n'-n\big)=0\,.\no
\eea

\noi Consequently, the infinite sum simply yields $D_{\rm inter}=2\cdot\Delta/\pi$. $D$ retains a universal value which can be associated with the quantum metric of the zeroth pseudo-Landau levels, with the former given by the definition: $g_{\varepsilon_\alpha}(q)=\sum_{\beta\neq\alpha}\big|\big<\partial_q u_\alpha(q)\big|u_\beta(q)\big>\big|^2\equiv
\sum_{\beta\neq\alpha}\big|\big<u_\alpha(q)\big|\partial_q u_\beta(q)\big>\big|^2$. For strained graphene hybrids, we have a contribution only from $(n,\sigma)=(0,-1)$ and find: $g_{\varepsilon_{n=0,\sigma=-1}}(q)=\sum_{\sigma'=\pm1}
\big|\big<u_{n=0,\sigma=-1}(q)\big|\partial_q u_{n=1,\sigma'}(q)\big>\big|^2=\ell_{\cal B}^2/2$.

\begin{figure}[b!]
\begin{center}
\includegraphics[width=\textwidth]{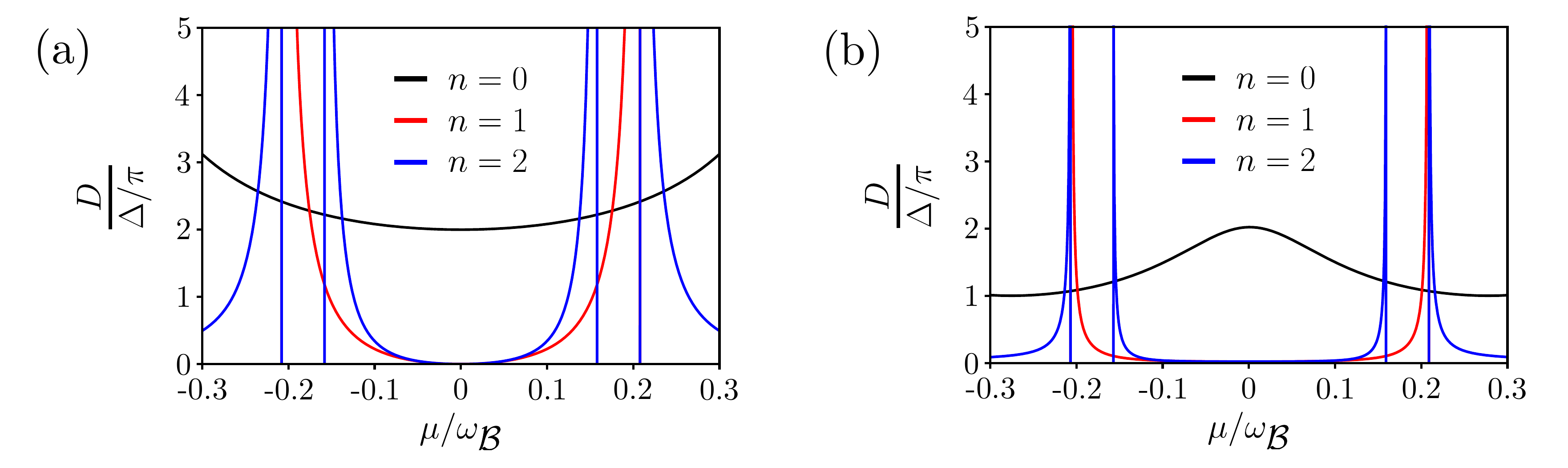}
\end{center}
\caption{Dependence of the superfluid stiffness with the chemical potential for the two limits of $\Delta$, (a) $\Delta/\omega_{\cal B}=10$ and (b) $\Delta/\omega_{\cal B}=0.1$, where only the contributions of the $n=0$, $1$, and $2$ levels are considered. In the limit of zero chemical potential and away from the resonance, the contribution of the $n\geq 1$ levels tends to zero. In the two limits, the superfluid stiffness undergoes a divergence corresponding to resonances.}
\label{fig:FigureApp1}
\end{figure}

\subsection{Superfluid Stiffness away from Charge Neutrality}

\noi At this point it is important to evaluate the above contribution when a nonzero chemical potential and a nonzero Zeeman field are introduced. {\color{black} In our companion paper~\cite{WJAextended} we have examined the case of superconducting unstrained graphene in the Dirac regime. Specifically,
we showed that for a higher-order Dirac point with vorticity $s$, the superfluid stiffness is given as $D^{(s)}(B,\mu)=|s|D_{\rm cone}(B,\mu)$, i.e., is $|s|$ times larger than the respective stiffness of a single Dirac cone which carries vorticity of a single unit. The superfluid stiffness of a single Dirac cone for $B=0$ and $\mu\neq0$ had been previously calculated in Ref.~\onlinecite{KopninSoninPRB}. In the same work it was shown that for $B=0$ and $|\mu|\gg\Delta$, one finds $D_{\rm cone}(B=0,|\mu|\gg\Delta)=|\mu|/2\pi$. We now proceed with the case of strained graphene. We find that for a nonzero chemical potential $\mu$ all pLLs have now nonzero contributions to the superfluid stiffness. For the 0pLLs we obtain:
\begin{align}
D_{\lambda,0{\rm pLL}}^{\mu\neq0}=\frac{\Delta}{\pi}\frac{1}{\sqrt{1+\big(\mu/\Delta\big)^2}}\frac{1}{1-\left(\frac{\mu}{\omega_{\cal B}/2}\right)^2}\,.\no
\end{align}

\noi Besides a renormalized $\Delta$, an additional factor emerges which diverges for $|\mu|=\omega_{\cal B}/2$. Thus, the stiffness can be strongly enhanced by tuning the system to this resonance. In fact, such re\-so\-nan\-ces appear for all pLLs, since we find:
}

\bea
D_{n=1,\sigma}^{\mu\neq0}&=&
2\cdot\frac{\Delta}{\pi}\frac{\Delta}{\sqrt{\big(1-\sigma\mu/\omega_{\cal B}\big)^2\omega_{\cal B}^2+\Delta^2}}\left\{\Theta\left[B+\sqrt{\big(1-\sigma\mu/\omega_{\cal B}\big)^2\omega_{\cal B}^2+\Delta^2}\right]-\Theta\left[B-\sqrt{\big(1-\sigma\mu/\omega_{\cal B}\big)^2\omega_{\cal B}^2+\Delta^2}\right]\right\}\cdot\no\\
&&\frac{1}{2}\Bigg\{-\frac{1}{1-\frac{\mu}{(\varepsilon_{0,-1}+\varepsilon_{1,\sigma})/2}}+\frac{1}{2}\sum_{\sigma'}\frac{1}{1-\frac{\mu}{(\varepsilon_{2,\sigma'}+\varepsilon_{1,\sigma})/2}}\Bigg\}\,.\no
\eea

\noi From the above, we observe that there exist resonances which are obtained by suitably tuning the chemical potential. Therefore, one can enhance the supercurrent drastically by approaching these. These resonances relate to the average energies of two pseudo-Landau levels in the absence of the chemical potential. Hence, this is an interesting feature that could be probed experimentally. Note also that as $\mu\rightarrow0$ we have $D_{n=1,\sigma}\rightarrow0$ for all values of $B$. Let us complete this analysis by considering the levels with $n\geq2$.
\begin{align}
D_{n\geq2,\sigma}^{\mu\neq0}=
2\cdot\frac{\Delta}{\pi}\frac{\Delta}{E_{n,\sigma}^{\mu\neq0}}\frac{1}{4}\sum_{\sigma'}\left[\frac{1}{1-\frac{\mu}{\big(\varepsilon_{n+1,\sigma'}+\varepsilon_{n,\sigma}\big)/2}}-\frac{1}{1-\frac{\mu}{\big(\varepsilon_{n-1,\sigma'}+\varepsilon_{n,\sigma}\big)/2}}\right]\Big[\Theta(B+E_{n,\sigma}^{\mu\neq0})-\Theta(B-E_{n,\sigma}^{\mu\neq0})\Big]\,.\no
\end{align}

\noi From the above we observe once again resonances for successive pseudo-Landau levels, since the resonance condition reads $\mu=\big(\varepsilon_{n\pm1,\sigma'}+\varepsilon_{n,\sigma}\big)/2$. Even more, taking the limit $\mu\rightarrow0$ yields that the contribution of the $n\geq1$ levels is exactly zero in this limit. In Fig.~\ref{fig:FigureApp1} we show numerical results.

\section{D. Strained Graphene Hybrids - Numerics}

We have performed additional calculations for the superfluid stiffness of both armchair and zigzag strained graphene nanoribbons. For all the numerics (including those of the main text) we have replaced the delta function appearing in Eq.~\eqref{eq:Dintra} by a Lorentzian with width $\Gamma=0.001$.

In Fig.~\ref{fig:FigureApp2} we present results for the same parameters as the ones presented in Fig.~2 of the main text, but for a zigzag strained graphene nanoribbon. Note that, in the zigzag case, we adopted a convention such that the pseudo-vector potential has only a nonzero $y$ component, which is generated via considering tensile strain and increases linearly in the $x$ direction. The latter also leads to an anisotropic {\color{black}Dirac velocity}. As a result, the pLLs become now weakly dispersive. This is in contrast to the armchair case, where the strain does not affect the isotropic character of the {\color{black}Dirac velocity}, thus rendering the pLLs entirely flat. We find that the quantization at $B=0$ is already spoiled due to contributions originating from additional edge flat bands that appear for zigzag terminations. The extra edge modes appear even without the presence of strain. Moreover, the drop of the superfluid stiffness across $|B|=\Delta$ is far from being quantized, thus implying that armchair graphene nanoribbons are more suitable for observing the topological quantization phenomena in the superfluid stiffness.

\begin{figure}[b!]
\centering
\includegraphics[width=\textwidth]{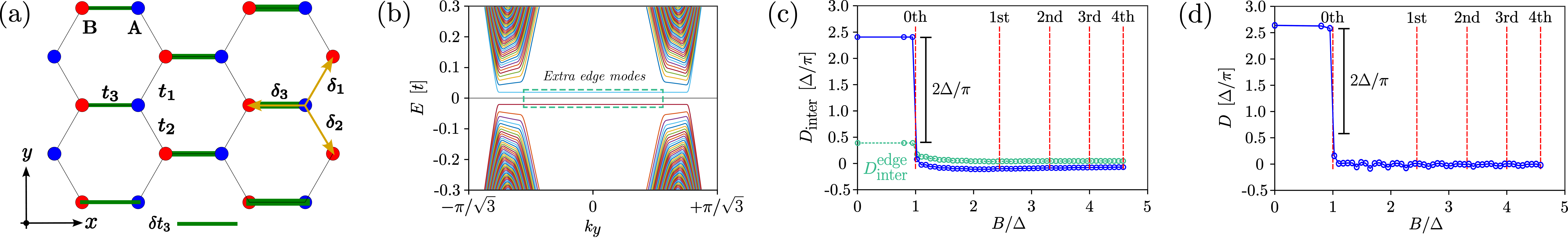}
\caption{(a) Schematic illustration of the model adopted for a zigzag graphene nanoribbon (GNR) under the influence of nonuniform strain. The strain is applied in a uniaxial manner so that the $y$ component of the pseudo-vector potential is nonzero and increasing with increasing $x$, as indicated by the thickness of the green lines. (b) Electronic band structure of (a) as a function of the conserved wave number $k_y$. We first numerically obtain the spectrum for an armchair GNR of width $W=601\times3a_0$ and $\omega_{\cal B}\simeq0.045$. Subsequently we add a conventional superconducting gap $\Delta=0.02$. Besides pLLs, additional edge flat bands emerge which are typical for zigzag GNRs and connect the two bulk Dirac cones. (c) and (d) correspond to the numerical results for the interband and total superfluid stiffness as a function of an applied inplane Zeeman field $B$ for the same values discussed in (b). In (c) we also show the contribution of the additional edge modes, which spoils the expected quantization of the interband term for $B=0$. The dashed vertical red lines indicate the energies for the pLLs in the presence of the nonzero $\Delta$. Due to these extra edge modes the quantization is more difficult to see in zigzag compared to armchair GNRs. Note that all energies are expressed in units of $t$, i.e., the nearest neighbor hopping in the absence of strain.}
\label{fig:FigureApp2}
\end{figure}
\begin{figure}[b!]
\centering
\includegraphics[width=\textwidth]{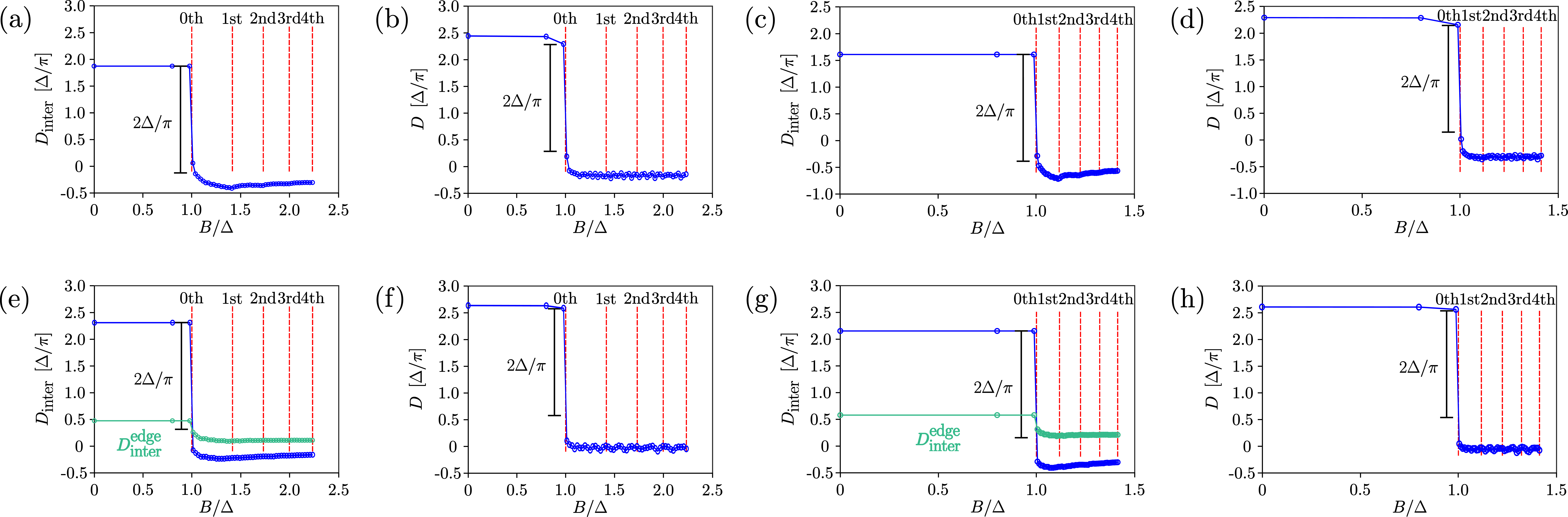}
\caption{Numerical results for the interband and total superfluid stiffness as a function of an applied inplane Zeeman field $B$ for armchair [(a)-(d)] and zigzag [(e)-(h)] GNRs for the same strain strength $\omega_{\cal B}\simeq0.045$. The panels (a), (b), (e) and (f) are obtained for $\Delta=\omega_{\cal B}$, while the panels (c), (d), (g) and (h) are obtained for $\Delta=2\omega_{\cal B}$. Note that all energies are expressed in units of $t$, i.e., the nearest neighbor hopping in the absence of strain.}
\label{fig:FigureApp3}
\end{figure}

We have also carried out studies for further values of the parameters $\Delta$ and $\omega_{\cal B}$. These are presented in Fig.~\ref{fig:FigureApp3}. From these results, we first find that increasing $\Delta/\omega_{\cal B}$ spoils the quantization of the geometric contribution to the superfluid stiffness. In spite of this fact, we find that the drop of the superfluid stiffness across $|B|=\Delta$ remains approximately quantized for armchair nanoribbons. In contrast, the extra edge flat bands appearing for zigzag nanoribbons contribute with a significant amount to the stiffness and introduce substantial deviations from the quantized value.

\section{E. Josephson Quantum Capacitance for Strained Graphene Hybrids}

To alternatively observe the underlying topological properties of bulk superconducting graphene in the hybrids under discussion, we consider that the superconducting gap picks up a time-dependent phase $\phi(t)$. This is possible by contacting the graphene hybrid of interest to a conventional superconductor through a dielectric which suppressess the Josephson coupling of the junction. $\phi$ corresponds to the phase difference appearing in the junction, while the junction sees a voltage bias $V$. Note that the gauge invariant electrostatic potential becomes $V\rightarrow V-\partial_t\phi/2$ (where we set $e=\hbar=1$). The above coupling naturally leads to the generation of an excess charge density for a nonzero $\partial_t\phi$. When $\partial_t\phi$ is constant, the above phenomenon leads to an additional contribution to the capacitive energy per area of the Josephson junction, which reads as $E_{\rm JJ}=-c_{\rm JJ}\big(V-\partial_t\phi/2\big)^2/2$. Here, $c_{\rm JJ}$ denotes the total capacitance per area which includes the so-called classical ($c_{C\ell}$) and quantum ($c_{\cal Q}$) parts. For the general class of quasi-one-dimensional systems of interest, the quantum ca\-pa\-ci\-tan\-ce $c_{\cal Q}$ is given by the respective charge susceptibility and takes the form:
\bea
c_{\cal Q}&=&-\int_{-\pi}^{+\pi}\frac{dq}{2\pi W}\int_{-\infty}^{+\infty}\frac{d\epsilon}{2\pi}{\rm Tr}\big[\tau_3\hat{G}(\epsilon,q)\tau_3\hat{G}(\epsilon,q)\big]\no\\
&=&-\int_{-\pi}^{+\pi}\frac{dq}{2\pi W}\int_{-\infty}^{+\infty}\frac{d\epsilon}{2\pi}{\rm Tr}\left[\tau_3\frac{i(\epsilon-iB)+\hat{h}(q)\tau_3+\Delta\tau_1}{(\epsilon-iB)^2+\hat{h}^2(q)+\Delta^2}\tau_3\frac{i(\epsilon-iB)+\hat{h}(q)\tau_3+\Delta\tau_1}{(\epsilon-iB)^2+\hat{h}^2(q)+\Delta^2}\right]\no\\
&=&\int_{-\pi}^{+\pi}\frac{dq}{2\pi W}\int_{-\infty}^{+\infty}\sum_n\frac{\Delta^2}{E_n^3(q)}\Big\{\Theta\big[B+E_n(q)\big]-\Theta\big[B-E_n(q)\big]\Big\}\,,\no
\eea

\noi where $n$ labels the eigenstates of $\hat{h}(q)$. Note that the above was derived under the assumption of an insulating system, i.e., the Bogoliubov-Fermi level set by $B$ does not cross any band. For the case of strained graphene hybrids harbouring pLLs, the quantum ca\-pa\-ci\-tan\-ce $c_{\cal Q}$ stemming from the Dirac cones of the two valleys is given by:

\begin{align}
c_{\cal Q}=2\cdot\frac{1}{2\pi\ell_{\cal B}^2}\sum_{(n,\sigma)}\frac{\Delta^2}{E_{n,\sigma}^3}\left[\Theta\big(B+E_{n,\sigma}\big)-\Theta\big(B-E_{n,\sigma}\big)\right].\no
\end{align}

\noi We focus on the $B=\mu=0$ case. The sum can be carried out exactly and yields:
\bea
c_{\cal Q}^{\mu=B=0}&=&2\cdot\frac{\Delta}{2\pi(\omega_{\cal B}\ell_{\cal B})^2}\sum_{(n,\sigma)}\frac{\Delta/\omega_{\cal B}}{\sqrt{n+\big(\Delta/\omega_{\cal B}\big)^2}^3}=
2\cdot\frac{\Delta}{4\pi\upsilon_D^2}\left\{2\left[\sum_n\frac{\Delta/\omega_{\cal B}}{\sqrt{n+\big(\Delta/\omega_{\cal B}\big)^2}^3}\right]-\left(\frac{\omega_{\cal B}}{\Delta}\right)^2\right\}\no\\
&=&
2\cdot\frac{\Delta}{\pi\upsilon_D^2}\left\{\left(\frac{\Delta}{2\omega_{\cal B}}\right)\zeta\left[\nicefrac{3}{2},\left(\Delta/\omega_{\cal B}\right)^2\right]-\left(\frac{\omega_{\cal B}}{2\Delta}\right)^2\right\},\no
\eea

\noi where $\zeta$ is the Hurwitz zeta function. The above expression sees the following limiting behaviours. First, in the weak strain regime $\omega_{\cal B}\ll\Delta$ we find $c_{\cal Q}^{\mu=B=0}=2\cdot\Delta/\pi\upsilon_D^2$. In the antipodal limit $\omega_{\cal B}\gg\Delta$, we obtain: $c_{\cal Q}^{\mu=B=0}=2/(2\pi\ell_{\cal B}^2\Delta)$.
\end{widetext}

\end{document}